\documentclass[pra,twocolumn,longbibliography,superscriptaddress]{revtex4-2}

\usepackage[T1]{fontenc}
\usepackage[utf8]{inputenc}
\usepackage{color}
\usepackage{babel}
\usepackage{amsmath}
\usepackage{amssymb}
\usepackage{subfigure}
\usepackage{graphicx}
\usepackage{physics} 
\usepackage{dsfont}
\usepackage{hyperref}
\usepackage{svg}
\usepackage{epsfig}
\usepackage{ragged2e}
\usepackage[normalem]{ulem}
\usepackage{comment}
\usepackage{braket}

\begin{document}

\title{Gauge-invariant thermodynamics of a finite-time quantum Otto engine}

\author{Midana Baial Sambú}
\email{midanasambu2016@gmail.com}
\affiliation{QPequi Group, Institute of Physics, Federal University of Goi\'as, Goi\^ania, Goi\'as, 74.690-900, Brazil}

\author{Thiago R. de Oliveira}
\email{troliveira@id.uff.br}
\affiliation{Instituto de Física, Universidade Federal Fluminense, Niterói, RJ, 24210-346,  Brazil}

\author{Lucas C. C\'eleri\href{https://orcid.org/0000-0001-5120-8176}{\includegraphics[scale=0.05]{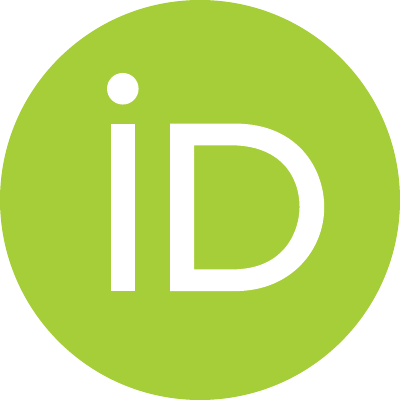}}}
\email{lucas@qpequi.com}
\affiliation{QPequi Group, Institute of Physics, Federal University of Goi\'as, Goi\^ania, Goi\'as, 74.690-900, Brazil}
\affiliation{Instituto de Física de São Carlos, Universidade de São Paulo, CP 369, 13560-970, São Carlos, SP, Brasil}

\begin{abstract}
We investigate a finite-time quantum Otto engine within the framework of gauge-invariant quantum thermodynamics, using the Lipkin-Meshkov-Glick model as the working medium. In this formulation, thermodynamic quantities are defined on equivalence classes of thermodynamically indistinguishable states, leading naturally to gauge-invariant notions of work, heat, entropy, and efficiency. We derive explicit expressions for these quantities and show that the usual work performed during the unitary strokes decomposes into an invariant contribution, associated with changes in the instantaneous energy spectrum and populations, and a coherent contribution arising from finite-time quantum coherences. This decomposition induces the corresponding splittings of engine efficiency and entropy production, providing a geometric interpretation of finite-time irreversibility and clarifying the role of coherence in work extraction. By analyzing the cycle's dependence on driving speed and system size, we identify the engine's operating region and investigate the influence of criticality. We show that crossing the critical region substantially reduces the parameter space in which the cycle operates as a heat engine. At the same time, whenever the engine condition remains satisfied, the discrepancy between the conventional and gauge-invariant descriptions becomes strongly suppressed, indicating that the gauge-invariant sector predominantly carries the extracted work. These results establish a direct connection between gauge-invariant thermodynamics, finite-time irreversibility, and work extraction in many-body quantum thermal machines.
\end{abstract}

\maketitle

\section{Introduction}
\label{sec:introduction}

Gauge principles occupy a central position in modern theoretical physics~\cite{Baez1994,konopleva1981gauge,faddeev1991gauge,nash2013topology,bleecker2013gauge}. Their importance stems from the observation that physically equivalent descriptions may contain redundant degrees of freedom that do not affect observable quantities. By identifying and systematically removing such redundancies, gauge theories provide a geometric framework that reveals the fundamental structures underlying a wide range of physical phenomena. Originally developed in the context of fundamental interactions, gauge ideas have subsequently found applications in various contexts~\cite{PhysRevLett.58.2051, Shapere_Wilczek_1989,Fradkin2013,GNNGauge2021,Vazquez2012, Katagiri2018,Borlenghi2016,polettini2012nonequilibrium}, where they serve as powerful tools for characterizing equivalence classes of descriptions, implementing coarse-graining procedures, and identifying the physically relevant information contained in complex systems.

From a different perspective, quantum thermodynamics seeks to extend the concepts and methods of thermodynamics to microscopic systems governed by the laws of quantum mechanics~\cite{Strasberg2022,Goold_2016,Alicki1979,Deffner2019}. During the last two decades, the field has experienced remarkable development. This progress has led to the formulation of quantum heat engines and refrigerators~\cite{Kieu2004,Kosloff2014}, the investigation of finite-time thermodynamic processes~\cite{Feldmann2003,Plastina2014,Campisi2011}, the study of nonequilibrium quantum dynamics~\cite{Talkner2007,Esposito2009,Campisi2011}, the exploration of the role played by genuinely quantum features such as coherence, entanglement, and quantum correlations~\cite{Lostaglio2016,Narasimhachar,Uzdin2015,Santos2019}, the development of thermodynamic uncertainty relations~\cite{Micadei2013}, and thermodynamics of relativistic~\cite{Basso2023,Basso2025,Costa2026,Moreira2025} and critical~\cite{Nascimento2024} quantum systems. (See Ref.~\cite{Potts2026} for an extensive list of references). Despite these advances, a fundamental conceptual challenge remains: quantum states generally contain substantially more information than is required for a thermodynamic description, or that is accessible to experimentalists, raising the question of how to formulate thermodynamics when only part of the microscopic information is operationally relevant. In particular, it is still not clear how to account for the coherences in the definition of work and heat, although some proposals interpret it as heat~\cite{Polkovnikov2008}, others consider it as work~\cite{Chen2018, Bertulio2020, Alipour2022, Ahmadi2023}. 

This observation motivates the recently developed gauge formulation of quantum thermodynamics~\cite{Celeri2024,Ferrari2025,Pernambuco2026a,Pernambuco2026b}. As in the classical world, the central idea is that quantum thermodynamics should not be formulated directly in the space of microscopic quantum states, but rather in the space of equivalence classes of thermodynamically indistinguishable states. Within this framework, the thermodynamic description emerges from a coarse-graining procedure associated with a thermodynamic gauge group. The states connected by transformations belonging to this group possess the same energetic description and are therefore regarded as equivalent from the thermodynamic point of view. As a consequence, thermodynamic quantities must be invariant under the action of the thermodynamic group, naturally leading to gauge-invariant notions of work, heat, and entropy. The novelty here is not that thermodynamics emerge from a coarse-graining procedure, but that this procedure can be unambiguously described as a gauge symmetry defined by the restricted access to information, thus providing a natural geometric theory for quantum thermodynamics~\cite{Pernambuco2026b}.

Beyond providing an alternative formulation of quantum thermodynamics, the gauge framework offers a geometric interpretation of the distinction between thermodynamically accessible and inaccessible information. It introduces new thermodynamic quantities such as coherent heat, gauge entropy, and invariant work. In particular, it allows one to separate contributions associated with populations in the instantaneous energy basis from those arising from quantum coherences. This separation decomposes thermodynamic quantities into invariant and coherence-related sectors. It provides a systematic way to identify the energetic and entropic contributions associated with information discarded by thermodynamic coarse-graining. From this perspective, the gauge formalism supplies a natural language for investigating the role of restricted information in nonequilibrium thermodynamics and for clarifying the physical significance of coherence-generated effects in driven quantum systems.

Quantum heat engines constitute a particularly suitable arena for investigating these questions. During finite-time operation, the driving protocols required to perform the work generally induce transitions between instantaneous energy levels and generate quantum coherences. These effects are responsible for departures from the quasistatic limit and are commonly associated with the phenomenon known as inner friction~\cite{Plastina2014}. Understanding how coherence generation, irreversibility, and work extraction are related remains an important problem in finite-time quantum thermodynamics. Moreover, when the working medium exhibits collective many-body behavior and critical phenomena, the interplay between coherence generation and the spectral properties of the system can lead to a particularly rich thermodynamic behavior~\cite{Campisi2016,Ma2017,Piccitto2022,Mukherjee2024}.

In this work, we investigate a finite-time quantum Otto engine with a working medium described by the Lipkin-Meshkov-Glick~\cite{Lipkin1965,Meshkov1965,Glick1965} model within the framework of gauge-invariant quantum thermodynamics. We derive explicit expressions for the gauge-invariant work, coherent heat, efficiency, and entropy production associated with the cycle, and we analyze their dependence on both the driving speed and the system size. The gauge decomposition reveals how finite-time contributions associated with quantum coherences modify the energetic and entropic balances of the engine and naturally leads to a separation between population and coherence sectors. We further show that this decomposition provides a geometric interpretation of finite-time irreversibility and suggests a natural connection between invariant work, restricted information, and work extraction from the thermodynamically accessible sector of the quantum state. Finally, by exploring different driving regimes of the Lipkin-Meshkov-Glick model, we investigate how criticality influences both the engine's operational region and the relative importance of coherent contributions to its performance.

The paper is organized as follows. In Sec.~\ref{sec:gauge}, we review the gauge formulation of quantum thermodynamics and introduce the main concepts and quantities required throughout the manuscript. The central results of the paper are presented in Sec.~\ref{sec:otto}, where we analyze the finite-time quantum Otto cycle and derive the corresponding thermodynamic quantities within both the conventional and gauge-invariant frameworks. Finally, Sec.~\ref{sec:conclusions} summarizes the main results and discusses their physical implications, highlighting several open questions and possible directions for future research.

\section{Gauge-invariant quantum thermodynamics}
\label{sec:gauge}

The gauge formulation of quantum thermodynamics is based on the observation that the microscopic quantum state generally contains more information than is thermodynamically accessible. Ordinary thermodynamics relies on a restricted description in which only a subset of observables is operationally relevant, while the remaining microscopic details are effectively removed by a coarse-graining procedure. In the quantum framework proposed in Ref.~\cite{Celeri2024} and further developed in Refs.~\cite{Ferrari2025,Pernambuco2026a,Pernambuco2026b}, this coarse-graining is implemented through a symmetry principle: thermodynamic quantities must be invariant under transformations that preserve the energetic description of the system~\cite{note1}. In this sense, gauge invariance identifies different microscopic descriptions that are thermodynamically indistinguishable.

Let $\rho_t$ be the state of a finite-dimensional quantum system with Hamiltonian $H_t$. The internal energy is given by $U[\rho_t]=\Tr(\rho_t H_t)$. A thermodynamic gauge transformation is defined as a unitary transformation $V_t$ that preserves the mean energy
\begin{equation}
U[V_t\rho_tV_t^\dagger]=U[\rho_t].
\end{equation}
Since this condition must be met for every density matrix, the allowed transformations satisfy $[V_t,H_t]=0$. Writing the Hamiltonian in its instantaneous spectral decomposition,
\begin{equation}
H_t=u_t h_t u_t^\dagger, \qquad h_t=\bigoplus_{k=1}^{p_t}\lambda_t^k\mathbb{I}_{n_t^k},
\end{equation}
where $\lambda_t^k$ are the distinct energy eigenvalues and $n_t^k$ their degeneracies, the gauge transformations take the form
\begin{equation}
V_t = u_t \left(\bigoplus_{k=1}^{p_t}v_t^k\right)u_t^\dagger, \qquad v_t^k\in\mathcal{U}(n_t^k).
\end{equation}
The thermodynamic group is therefore
\begin{equation}
\mathcal{G}_{\mathrm T} \simeq \mathcal{U}(n_t^1)\times\mathcal{U}(n_t^2)\times\cdots\times\mathcal{U}(n_t^{p_t}),
\end{equation}
which identifies the transformations that leave the system's mean energy unchanged. The role of this group is not to describe physical dynamics, but rather to characterize redundant information from the thermodynamic point of view~\cite{Celeri2024,Ferrari2025,Pernambuco2026a,Pernambuco2026b}.

The usual formulation of the first law decomposes the variation of the internal energy as $\Delta U=W_{\mathrm u}[\rho]+Q_{\mathrm u}[\rho]$, where the work and heat functionals are given by~\cite{Alicki1979}
\begin{equation}
W_{\mathrm u}[\rho_t] = \int_0^\tau \dd t\, \Tr(\rho_t\dot H_t),
\label{eq:usual_work}
\end{equation}
and
\begin{equation}
Q_{\mathrm u}[\rho_t] = \int_0^\tau \dd t \Tr(\dot\rho_tH_t),
\label{eq:usual_heat}
\end{equation}
respectively. Here, $\tau$ denotes the duration of the thermodynamic process.

In the gauge formulation, thermodynamic quantities are obtained by requiring invariance under the thermodynamic group. Technically, this is achieved by averaging the usual quantities over $\mathcal{G}_{\mathrm T}$. Applying this procedure to work yields the gauge-invariant work~\cite{Celeri2024}
\begin{equation}
W_{\mathrm{inv}}[\rho_t] = \int_0^\tau \dd t \Tr\left( \rho_t u_t\dot h_tu_t^\dagger\right).
\label{eq:invariant_work}
\end{equation}
This expression shows that the invariant work is determined solely by changes in the instantaneous energy eigenvalues. Equivalently, it is the work assigned by a thermodynamic description that is insensitive to the information removed by the gauge coarse-graining. In this sense, $W_{\mathrm{inv}}$ is not, in general, the total mechanical work performed during the process. Rather, it represents the contribution to the work that survives after discarding the information inaccessible within the restricted thermodynamic description.

The invariant heat is then fixed by the first law $\Delta U=W_{\mathrm{inv}}[\rho]+Q_{\mathrm{inv}}[\rho]$. It can be written as $Q_{\mathrm{inv}}[\rho_t] = Q_{\mathrm u}[\rho_t] + Q_{\mathrm c}[\rho_t]$, where
\begin{equation}
Q_{\mathrm c}[\rho_t] = \int_0^\tau \dd t\, \Tr\left( \rho_t\dot u_th_tu_t^\dagger + \rho_t u_th_t\dot u_t^\dagger \right)
\label{eq:coherent_heat}
\end{equation}
is the coherent heat. Consequently, $W_{\mathrm{inv}}[\rho_t] = W_{\mathrm u}[\rho_t]-Q_{\mathrm c}[\rho_t]$.

The coherent heat quantifies the energetic contribution associated with the changes of the instantaneous energy basis and, consequently, with the coherences generated in that basis. Therefore, it represents the difference between the total mechanical work and the work inferred from the gauge-invariant sector. From the perspective of the restricted observer, $Q_{\mathrm c}$ accounts for the energetic correction arising from the information discarded by the thermodynamic coarse graining. Since it is not associated with the expectation value of a positive operator, its sign is not fixed by the formalism.

The same gauge principle also gives rise to a thermodynamic entropy. Let $\rho_t^E=u_t^\dagger\rho_tu_t$ be the density matrix written in the instantaneous energy basis. Averaging over the thermodynamic group produces the block-diagonal state
\begin{equation}
\rho_{dd}^E(t) = \bigoplus_{k=1}^{p_t} \frac{\Tr[\rho_{n_t^k}^E(t)]}{n_t^k} \mathbb{I}_{n_t^k},
\end{equation}
where $\rho_{n_t^k}^E(t)$ denotes the projection of $\rho_t^E$ onto the eigenspace associated with the eigenvalue $\lambda_t^k$. Taking the von Neumann entropy as the underlying information measure, the gauge-invariant entropy is defined as
\begin{equation}
S_{\mathcal{G}_{\mathrm{T}}}[\rho_t] = S[\rho_{dd}^E(t)],
\end{equation}
with $S[\rho]=-\Tr(\rho\log\rho)$.

For practical applications, it is convenient to express this quantity as
\begin{equation}
S_{\mathcal G_T}[\rho_t] = S_{\mathrm d}[\rho_t] + S_\Gamma[f_t],
\label{eq:invariant_entropy}
\end{equation}
where $S_{\mathrm d}[\rho_t] = S[\rho_{\mathrm{diag}}^E(t)]$ is the diagonal entropy~\cite{Jaynes1957} in the instantaneous energy basis, while $S_\Gamma[f_t] = -\Tr(f_t\log|f_t|)$ accounts for the contribution arising from degeneracies of the energy spectrum. The explicit form of the block-diagonal object $f_t$ is a block-diagonal object that encodes the difference between the Haar-averaged state and the purely diagonal state. It is explicitly given in Ref.~\cite{Ferrari2025}.

The two contributions entering Eq.~\eqref{eq:invariant_entropy} have distinct physical interpretations. The diagonal entropy quantifies the uncertainty associated with the energy populations after removing coherences in the instantaneous energy basis. The correction $S_\Gamma$ captures the additional contribution generated by the degeneracy structure of the spectrum. Therefore, $S_{\mathcal G_T}[\rho_t]$ is a thermodynamic entropy associated with the pair $(\rho_t,H_t)$ and with the coarse graining induced by the thermodynamic group. Consequently, while the von Neumann entropy remains constant during unitary dynamics, the gauge-invariant entropy may vary whenever the instantaneous energy basis or the degeneracy structure of the Hamiltonian changes in time.

\section{The quantum Otto engine}
\label{sec:otto}

In this section, we investigate the performance of a quantum Otto engine within the framework of the gauge-invariant thermodynamics introduced above. We begin by introducing the working medium of the engine, namely the Lipkin-Meshkov-Glick (LMG) model~\cite{Lipkin1965,Meshkov1965,Glick1965}, described by the Hamiltonian
\begin{equation}
    H(t) = -\frac{\gamma_x}{2j}J_x^2-\delta_tJ_z,
    \label{eq:LMGmodel}
\end{equation}
where $J_\alpha=\sum_i\sigma_\alpha^i$ denotes the collective angular momentum operator along the direction $\alpha=\{x,y,z\}$ for a system of $N=2j$ spin-$1/2$ particles, and $\sigma_\alpha^i$ is the Pauli operator acting on the $i$-th spin. The parameters $\gamma_x>0$ and $\delta_t\geq0$ correspond to the interaction strength and the transverse magnetic field, respectively. We allow the magnetic field to explicitly depend on time to implement the driving protocol of the heat engine. Specifically, we consider
\begin{equation}
    \delta_t=\delta\pm\delta_0\tanh(\alpha t),
    \label{magneticfiledtime}
\end{equation}
where $\delta_0=\delta_2-\delta_1$ is the difference between the final and initial values of the field during a given stroke. The parameter $\delta$ is chosen as $\delta_1$ or $\delta_2$, depending on the stage of the cycle, while $\alpha$ controls the speed of the driving protocol. From here on we employ the short-hand notation $H_{\delta_i} \equiv H(\delta_t = \delta_i)$, for $i=1,2$.

For a time-independent transverse field, the LMG model exhibits a second-order ground-state quantum phase transition in the thermodynamic limit. The competition between the collective interaction and the magnetic field gives rise to two distinct phases separated by the critical point $\delta_c=\gamma_x$. For $\delta>\gamma_x$, the ground state is polarized along the field direction and preserves the parity symmetry of the Hamiltonian. In contrast, for $\delta<\gamma_x$, the interaction term dominates, and the system enters a ferromagnetic phase characterized by finite transverse magnetization and spontaneous symmetry breaking. The transition is accompanied by the closing of the excitation gap in the thermodynamic limit~\cite{Lipkin1965,Meshkov1965,Glick1965}.

The critical behavior extends beyond the ground state, as evidenced by an excited-state quantum phase transition. For $\delta<\gamma_x$, the spectrum exhibits a singularity at the critical energy $E_c=-j\delta$, which corresponds, in the semiclassical picture, to the separatrix of the underlying classical dynamics. At this energy, the density of states diverges in the thermodynamic limit, and the structure of the eigenstates changes abruptly, separating regions of the spectrum with qualitatively different properties~\cite{Corps2022}.

The LMG model also exhibits dynamical quantum phase transitions following sudden quenches across the critical point, signaled by nonanalyticities in the Loschmidt rate function at critical times in the thermodynamic limit~\cite{Heyl2014}. Although the finite-time protocols considered in the present work interpolate between smooth transformations and increasingly quench-like drivings, we do not explicitly investigate dynamical quantum phase transitions here.

We now consider the LMG model as the working medium of a quantum Otto cycle. The cycle consists of four strokes: two isochoric processes, during which the system exchanges heat with thermal reservoirs and evolves non-unitarily, and two unitary strokes, during which the system is isolated from the environment and evolves under the action of a time-dependent Hamiltonian.

The protocol starts at time $t_0$, with the working medium prepared in thermal equilibrium with a hot reservoir at temperature $T_h$. At this stage, the magnetic field is fixed at $\delta_{t=0}=\delta_1$, and the initial state is given by
\begin{equation}
    \rho_A = \frac{1}{Z_{\delta_1}} e^{-H_{\delta_1}/T_h},
\end{equation}
where $Z_{\delta_1} = \Tr\left(e^{-H_{\delta_1}/T_h}\right) $ is the corresponding partition function.

During the first stroke, the system is decoupled from the hot reservoir and undergoes a unitary expansion process. The state evolves according to $\rho_B = U_t^{\mathrm{exp}}\rho_A(U_t^{\mathrm{exp}})^\dagger$, where
\begin{equation}
    U_t = \vec{\mathrm T}\exp\left[-i\int \dd t\,H_t\right]
\end{equation}
is the time-evolution operator and $\vec{\mathrm T}$ denotes the time-ordering operator. During this process, the control parameter varies according to $\delta_t=\delta_1+\delta_0\tanh(\alpha t)$, driving the Hamiltonian from $H_{\delta_1}$ to $H_{\delta_2}$ over a finite time interval $\tau$. The work performed during this stroke is denoted by $W_{\mathrm{id}}^{\mathrm{exp}}$, where the label $\mathrm{id}$ refers to either the usual definition ($\mathrm u$) or the gauge-invariant one ($\mathrm{inv}$).

In the second stroke, the working medium is brought into contact with a cold reservoir at a temperature $T_c<T_h$. The resulting non-unitary evolution thermalizes the system to the Gibbs state associated with the Hamiltonian $H_{\delta_2}$,
\begin{equation}
    \rho_C = \frac{1}{Z_{\delta_2}} e^{-H_{\delta_2}/T_c},
\end{equation}
where $Z_{\delta_2} = \Tr\left(e^{-H_{\delta_2}/T_c}\right)$. The heat exchanged with the cold reservoir during this isochoric process is denoted by $Q_{\mathrm{id}}^{\mathrm{c}}$.

During the third stroke, the system is once again isolated from the environment and undergoes a unitary compression process. The state evolves from $\rho_C$ to $\rho_D = U_t^{\mathrm{comp}}\rho_C(U_t^{\mathrm{comp}})^\dagger$, while the control parameter varies according to $\delta_t=\delta_2-\delta_0\tanh(\alpha t)$, thus restoring the Hamiltonian from $H_{\delta_2}$ to $H_{\delta_1}$ during the same time interval $\tau$. The work associated with this stroke is denoted by $W_{\mathrm{id}}^{\mathrm{comp}}$, and the total work performed during the cycle is given by
\begin{equation}
    W_{\mathrm{id}}^{\mathrm{tot}} = W_{\mathrm{id}}^{\mathrm{exp}} + W_{\mathrm{id}}^{\mathrm{comp}}.
    \label{eq:tot_work}
\end{equation}

Finally, in the fourth stroke, the system is reconnected to the hot reservoir and undergoes a non-unitary evolution back to the initial thermal state $\rho_A$, thus closing the cycle. The heat exchanged with the hot reservoir during this process is denoted by $Q_{\mathrm{id}}^{h}$.

Having established the Otto cycle and the relevant thermodynamic quantities, we now investigate how the gauge decomposition modifies the energetic description of the engine and analyze its performance in terms of both the physical efficiency and its gauge-invariant and coherent contributions.

\subsection{Work}

We begin by observing that no work is performed during the isochoric strokes of the Otto cycle. Since the Hamiltonian remains constant during the thermalization processes, Eqs.~\eqref{eq:usual_work} and~\eqref{eq:invariant_work} immediately imply that $W_{\mathrm u}[\rho_t]=W_{\mathrm{inv}}[\rho_t]=0$. Therefore, it is sufficient to compute the work performed during the two unitary strokes, namely the expansion and compression processes.

The usual and gauge-invariant formulations generally assign different values to the work performed during the unitary strokes. The origin of this difference is that finite-time driving generates transitions between the instantaneous energy eigenstates and coherences in that basis, which contribute differently to the two thermodynamic descriptions. In what follows, we evaluate both quantities, starting from the usual formulation.

During the expansion and compression strokes, the dynamics are purely unitary. Consequently, Eq.~\eqref{eq:usual_heat} implies that $Q_{\mathrm u}[\rho_t]=0$, and the usual work reduces to the corresponding variation of the internal energy. For the expansion stroke, we obtain $W_{\mathrm u}^{\mathrm{exp}} = \Tr\left[H_{\delta_2}\rho_B - H_{\delta_1}\rho_A\right]$, while the compression stroke yields $W_{\mathrm u}^{\mathrm{comp}} = \Tr\left[H_{\delta_1}\rho_D - H_{\delta_2}\rho_C \right]$. Using the explicit form of the LMG Hamiltonian and defining $m_{x,\mu}^{2} = \Tr(J_x^2\rho_\mu)$ and $m_{z,\mu} = \Tr(J_z\rho_\mu)$, with $\mu=A,B,C,D$, the total work takes the form
\begin{eqnarray}
W_{\mathrm u}^{\mathrm{tot}} &=& -\frac{\gamma_x}{2j}\left[m_{x,B}^{2} - m_{x,C}^{2} + m_{x,D}^{2} - m_{x,A}^{2}\right] \nonumber\\
&-&\delta_2 \left(m_{z,B}-m_{z,C}\right) -\delta_1\left(m_{z,D}-m_{z,A}\right).
\label{eq:total_work_usual}
\end{eqnarray}

Let us now turn to the gauge formulation of quantum thermodynamics. In order to compute the invariant work defined in Eq.~\eqref{eq:invariant_work}, we consider the instantaneous energy eigenbasis of the Hamiltonian, $H_t = \sum_n \varepsilon_t^n \ket{\varepsilon_t^n}\bra{\varepsilon_t^n}$. In this basis, Eq.~\eqref{eq:invariant_work} can be written as
\begin{equation}
W_{\mathrm{inv}} = \int_0^\tau \dd t \sum_n p_t^n\, \dot{\varepsilon}_t^n,
\label{eq:Winv_populations}
\end{equation}
where $p_t^n= \bra{\varepsilon_t^n}\rho_t\ket{\varepsilon_t^n}$ denotes the instantaneous population of the energy level $\varepsilon_t^n$.

Using the Hellmann-Feynman theorem together with the LMG Hamiltonian, we find $\dot{\varepsilon}_t^n = -\dot{\delta}_t \bra{\varepsilon_t^n}J_z\ket{\varepsilon_t^n}$. Substituting this result into Eq.~\eqref{eq:Winv_populations}, the invariant work becomes
\begin{equation}
W_{\mathrm{inv}} = -\int_0^\tau \dd t\, \dot{\delta}_t m_z^{E}(t),
\label{eq:Winv_magnetization}
\end{equation}
where we introduced the magnetization projected onto the instantaneous energy basis,
\begin{equation}
m_z^{E}(t) = \sum_n p_t^n \bra{\varepsilon_t^n}J_z\ket{\varepsilon_t^n} = \Tr\left[
\rho_t^{E,\mathrm{diag}}J_z^{E}\right].
\label{eq:coherent_mag}
\end{equation}
Here, $\rho_t^{E,\mathrm{diag}} = \sum_n p_t^n \dyad{\varepsilon_t^n}$ is the diagonal part of the density matrix in the instantaneous energy basis and $J_z^{E} = u_t^\dagger J_z u_t$.

Finally, using the protocol defined in Eq.~\eqref{magneticfiledtime}, the total invariant work performed during the Otto cycle takes the form
\begin{equation}
W_{\mathrm{inv}}^{\mathrm{tot}} = \delta_0\alpha \int_0^\tau \dd t\,\sech^2(\alpha t)\left[m_{z,\mathrm{comp}}^{E}(t) - m_{z,\mathrm{exp}}^{E}(t)\right],
\label{eq:total_invariant_work}
\end{equation}
where the notation $m_{z,\mathrm{str}}^{E}(t)$ indicates that the density operator entering the definition of $m_z^{E}(t)$ must be evaluated using the state corresponding to the stroke $\mathrm{str}\in\{\mathrm{exp},\mathrm{comp}\}$.

\subsection{Heat}

We begin by considering the isochoric strokes, during which the system exchanges heat with the environment, and thermalization takes place. During these stages, the dynamics of the working medium is described by the master equation
\begin{equation}
\dot{\rho}_t=-i[H_t,\rho_t]+\mathcal{D}[\rho_t],
\label{eq:master}
\end{equation}
where the dissipative contribution $\mathcal{D}$ is typically assumed to have the Lindblad form. Throughout this work, we consider the weak-coupling regime between the system and the environment, implying that thermalization occurs over long timescales and entropy production during the relaxation process is negligible.

Since the Hamiltonian remains constant during the thermalization stages, its eigenbasis does not evolve in time and therefore $\dot{u}_t=0$. It then follows from Eq.~\eqref{eq:coherent_heat} that the coherent contribution to the heat vanishes during both isochoric processes. Consequently, the invariant heat reduces to the usual heat,
\begin{equation}
Q_{\mathrm{inv}}[\rho_t] = Q_{\mathrm u}[\rho_t] = \int_0^{\tau_{\mathrm{th}}}\dd t\,\Tr\left[H_t\mathcal{D}(\rho_t)\right],
\label{heatwithlindiblad}
\end{equation}
where $\tau_{\mathrm{th}}$ denotes the thermalization time. Although the formal expression is the same for both reservoirs, the actual amount of exchanged heat depends on the corresponding dissipative couplings, which are generally different and reflect, among other factors, the temperatures of the environments.

Moreover, as discussed previously, the work vanishes during isochoric strokes. Therefore, the heat exchanged during each thermalization process is completely determined by the variation of the internal energy.

For the second stroke, in which the working medium thermalizes with the cold reservoir, the exchanged heat is given by
\begin{equation}
Q_{\mathrm{inv}}^{\mathrm c} = Q_{\mathrm u}^{\mathrm c} = \Tr\left[H_{\delta_2}(\rho_C-\rho_B)\right],
\label{coldheat}
\end{equation}
Similarly, during the fourth stroke, corresponding to the thermalization with the hot reservoir, one obtains
\begin{equation}
Q_{\mathrm{inv}}^{\mathrm h} = Q_{\mathrm u}^{\mathrm h} = \Tr\left[H_{\delta_1}(\rho_A-\rho_D)\right],
\label{hotheat}
\end{equation}

Let us now turn to the unitary strokes. Since the dynamics during the expansion and compression processes is governed solely by the unitary part of Eq.~\eqref{eq:master}, Eq.~\eqref{eq:usual_heat} immediately implies that $Q_{\mathrm u}=0$ for both processes. However, because the Hamiltonian depends explicitly on time, the coherent contribution in Eq.~\eqref{eq:coherent_heat} does not necessarily vanish. In this case, the invariant heat is entirely determined by the coherent contribution, $Q_{\mathrm{inv}} = Q_{\mathrm c}$.

For the LMG model, the coherent heat can be expressed in a particularly simple form. Using Eq.~\eqref{eq:coherent_heat} together with the Hellmann-Feynman theorem, one finds
\begin{equation}
Q_{\mathrm c} = -\int_0^\tau \dd t\, \dot{\delta}_t\left[m_z(t)-m_z^{E}(t)\right],
\label{eq:coherent_heat_LMG}
\end{equation}
where $m_z(t) = \Tr(\rho_tJ_z)$ is the total magnetization and $m_z^{E}(t)$ is the energy-basis magnetization defined in Eq.~\eqref{eq:coherent_mag}. It is therefore useful to introduce the quantity $m_z^{\mathrm c}(t) = m_z(t)-m_z^{E}(t)$, which measures the difference between the full magnetization and its projection onto the instantaneous energy basis. Equation~\eqref{eq:coherent_heat_LMG} then becomes
\begin{equation}
Q_{\mathrm c} = -\int_0^\tau \dd t\,\dot{\delta}_t m_z^{\mathrm c}(t).
\end{equation}

Using the driving protocol defined in Eq.~\eqref{magneticfiledtime}, the total coherent heat generated during the two unitary strokes, $Q_{\mathrm c}^{\mathrm{tot}} = Q_{\mathrm c}^{\mathrm{exp}} + Q_{\mathrm c}^{\mathrm{comp}}$ is therefore
\begin{equation}
Q_{\mathrm c}^{\mathrm{tot}} =  \delta_0\alpha\int_0^\tau\dd t\,\sech^2(\alpha t)\left[m_{z,\mathrm{comp}}^{\mathrm c}(t)-m_{z,\mathrm{exp}}^{\mathrm c}(t)\right].
\label{eq:total_coherent_heat}
\end{equation}

The heat balance of the Otto cycle can thus be summarized as follows. During the isochoric strokes, the coherent contribution vanishes and the invariant and usual formulations coincide, $Q_{\mathrm{inv}}^{\mathrm c} = Q_{\mathrm u}^{\mathrm c}$ and $Q_{\mathrm{inv}}^{\mathrm h} = Q_{\mathrm u}^{\mathrm h}$. By contrast, during the unitary strokes the usual heat vanishes, $Q_{\mathrm u}^{\mathrm{exp}} = Q_{\mathrm u}^{\mathrm{comp}} = 0$, while the invariant heat is entirely determined by the coherent contribution, $Q_{\mathrm{inv}}^{\mathrm{exp}} = Q_{\mathrm c}^{\mathrm{exp}}$ and $Q_{\mathrm{inv}}^{\mathrm{comp}} = Q_{\mathrm c}^{\mathrm{comp}}$.

This completes the energetic characterization of the heat exchanged during the four stages of the quantum Otto cycle. Before considering the efficiency of the engine, we establish a relation between the coherent heat and the quantum coherence generated during any unitary process.

\subsection{Coherent heat and quantum coherences}

The previous subsection established that, during the unitary strokes of the Otto cycle, the invariant heat is entirely determined by the coherent contribution. It is therefore natural to ask whether coherent heat can be related to a quantitative measure of quantum coherence. In this subsection, we show that the magnitude of the coherent heat is bounded by the relative entropy of coherence, thereby establishing a direct connection between gauge-invariant thermodynamics and resource-theoretic quantifiers of coherence.

From Eq.~\eqref{eq:coherent_heat}, the coherent heat generated during a unitary stroke can be written as
\begin{equation}
Q_{\mathrm c} = -\int_0^\tau \dd t\,\dot{\delta}_t\Tr\left[\rho_t^{E,\mathrm{coh}}J_z^{E}(t)\right],
\label{eq:Qc_coh_trace}
\end{equation}
where $\rho_t^{E,\mathrm{coh}} = \rho_t^{E} - \rho_t^{E,\mathrm{diag}}$ denotes the off-diagonal contribution of the density matrix in the instantaneous energy basis. Taking the absolute value of this equation and using the triangle inequality, we obtain
\begin{equation}
\abs{Q_{\mathrm c}} \leq \int_0^\tau \dd t\, \abs{\dot{\delta}_t}
\abs{\Tr\left[\rho_t^{E,\mathrm{coh}}J_z^{E}(t)\right]}.
\end{equation}
Applying H\"{o}lder's inequality to the integrand yields
\begin{equation}
\abs{\Tr\left[\rho_t^{E,\mathrm{coh}}J_z^{E}(t)\right]} \leq\left\|\rho_t^{E,\mathrm{coh}}\right\|_1\left\|J_z^{E}(t)\right\|_\infty,
\end{equation}
where $\|\cdot\|_1$ and $\|\cdot\|_\infty$ denote the trace norm and the operator norm, respectively.

Since $J_z^{E}(t)=u_t^\dagger J_z u_t$ is related to $J_z$ by a unitary transformation, both operators possess the same spectrum and, therefore, the same operator norm, $\left\|J_z^{E}(t)\right\|_\infty =\|J_z\|_\infty=j$. Consequently,
\begin{equation}
\abs{Q_{\mathrm c}} \leq j \int_0^\tau\dd t\,\abs{\dot{\delta}_t} \left\| \rho_t^{E,\mathrm{coh}}\right\|_1.
\label{eq:Qc_trace_bound}
\end{equation}

To relate this expression to a coherence monotone, we consider the relative entropy of coherence~\cite{Baumgratz2014}
\begin{equation}
C_{\mathrm{rel}}(\rho_t) = D\left(\rho_t\big\|\rho_t^{E,\mathrm{diag}}\right),
\end{equation}
where $\rho_t^{E,\mathrm{diag}}$ is the dephased state in the instantaneous energy basis and $D(\rho\|\sigma) = \Tr\left[\rho\left(\log\rho-\log\sigma\right)\right]$ denotes the quantum relative entropy.

Since $\rho_t^{E,\mathrm{coh}} = \rho_t - \rho_t^{E,\mathrm{diag}}$,  Pinsker's inequality implies
\begin{equation}
\left\| \rho_t^{E,\mathrm{coh}} \right\|_1 = \left\|\rho_t-\rho_t^{E,\mathrm{diag}}\right\|_1\leq\sqrt{2D\left(\rho_t\big\|\rho_t^{E,\mathrm{diag}}\right)},
\end{equation}
thus leading to
\begin{equation}
\abs{Q_{\mathrm c}} \leq j\sqrt{2}\int_0^\tau\dd t\,\abs{\dot{\delta}_t}\sqrt{C_{\mathrm{rel}}(\rho_t)}.
\label{eq:Qc_coherence_bound}
\end{equation}

Equation~\eqref{eq:Qc_coherence_bound} establishes an upper bound on the coherent heat in terms of a well-established resource-theoretic quantifier of coherence. It shows that the magnitude of the energetic contribution associated with thermodynamically inaccessible coherences is constrained by the amount of coherence present in the instantaneous energy basis. At the same time, the coherent heat is not determined solely by the coherence monotone itself. While $C_{\mathrm{rel}}$ quantifies the distinguishability between the actual state and its dephased counterpart, the coherent heat also depends on the driving protocol through the factor $\abs{\dot{\delta}_t}$. Consequently, two processes exhibiting the same amount of coherence may generate different coherent heats depending on how the external driving couples to these coherences.

Although derived here for the LMG model, the argument relies only on Hölder's and Pinsker's inequalities and therefore extends straightforwardly to arbitrary driven quantum systems. Starting directly from Eq.~\eqref{eq:coherent_heat}, one finds the general bound
\begin{equation}
\abs{Q_{\mathrm c}} \leq \sqrt{2} \int_0^\tau\dd t\,\left\|u_t^\dagger\dot{H}_t u_t - \dot{h}_t \right\|_\infty\sqrt{C_{\mathrm{rel}}(\rho_t)}.
\end{equation}
This result establishes a direct connection between coherent heat and quantum coherence. Rather than identifying the coherent heat with a coherence monotone, it shows that the energetic contribution associated with thermodynamically inaccessible coherences is constrained by a resource-theoretic measure of coherence. In this sense, coherence bounds the maximal magnitude of the coherent correction appearing in the gauge decomposition of the first law.

\subsection{Efficiency}

We adopt the convention that energy flowing out of the working medium is positive while energy entering the system is negative. Within this convention, the Otto cycle operates as a heat engine whenever the net work extracted from the working medium and the heat absorbed from the hot reservoir satisfy $W_{\mathrm u}^{\mathrm{tot}}>0$ and $Q_{\mathrm u}^{h}>0$. The usual efficiency of the Otto cycle is therefore defined as
\begin{equation}
    \eta_{\mathrm O} = \frac{W_{\mathrm u}^{\mathrm{tot}}}{Q_{\mathrm u}^{h}} = \frac{W_{\mathrm u}^{\mathrm{tot}}}{Q_{\mathrm{inv}}^{h}},
    \label{eq:phys_efficiency}
\end{equation}
where we used the fact that $Q_{\mathrm{inv}}^{h}=Q_{\mathrm u}^{h}$, as established previously. Since the total work admits the decomposition $W_{\mathrm u}^{\mathrm{tot}} = W_{\mathrm{inv}}^{\mathrm{tot}} + Q_{\mathrm c}^{\mathrm{tot}}$, with $Q_{\mathrm c}^{\mathrm{tot}} = Q_{\mathrm c}^{\mathrm{exp}} + Q_{\mathrm c}^{\mathrm{comp}}$, the efficiency can be written as
\begin{equation}
    \eta_{\mathrm O} = \eta_{\mathrm{inv}} + \eta_{\mathrm c},
\end{equation}
where
\begin{equation}
    \eta_{\mathrm{inv}} = \frac{W_{\mathrm{inv}}^{\mathrm{tot}}}{Q_{\mathrm u}^{h}}, \qquad \eta_{\mathrm c} = \frac{Q_{\mathrm c}^{\mathrm{tot}}}{Q_{\mathrm u}^{h}}.
    \label{eq:split_efficiency}
\end{equation}

We are interested in the finite-time cycle for distinct system sizes and the speed of the unitary processes. In order to compare different driving rates on equal footing, we consider a family of cycles that connect the same initial and final Hamiltonians for all values of $\alpha$. Since the protocol defined in Eq.~\eqref{magneticfiledtime} reaches its asymptotic value only in the limit $t\rightarrow\infty$, the duration of unitary strokes is not fixed independently. Instead, for each value of $\alpha$, we determine a stroke duration $\tau_\alpha$ requiring the expansion protocol to reach the target value $\delta_2$ up to a prescribed numerical tolerance $\epsilon$. Explicitly, we impose $\tanh(\alpha\tau_\alpha) = 1-\epsilon$, which gives
\begin{equation}
\tau_\alpha = \frac{\operatorname{arctanh}(1-\epsilon)}{\alpha}.
\end{equation}
The same duration $\tau_\alpha$ is used for the compression stroke. Due to the protocol's symmetry, the compression process returns to $\delta_1$ within the same numerical tolerance. Consequently, each value of $\alpha$ defines a complete Otto cycle with the same pair of boundary Hamiltonians, $H_{\delta_1}$ and $H_{\delta_2}$, differing only in the time required to implement the unitary transformations. This construction ensures that changes in the thermodynamic quantities are due to the nonadiabatic character of the driving, rather than to changes in the endpoints of the cycle.

\begin{figure}
    \centering
    \includegraphics[width=1\linewidth]{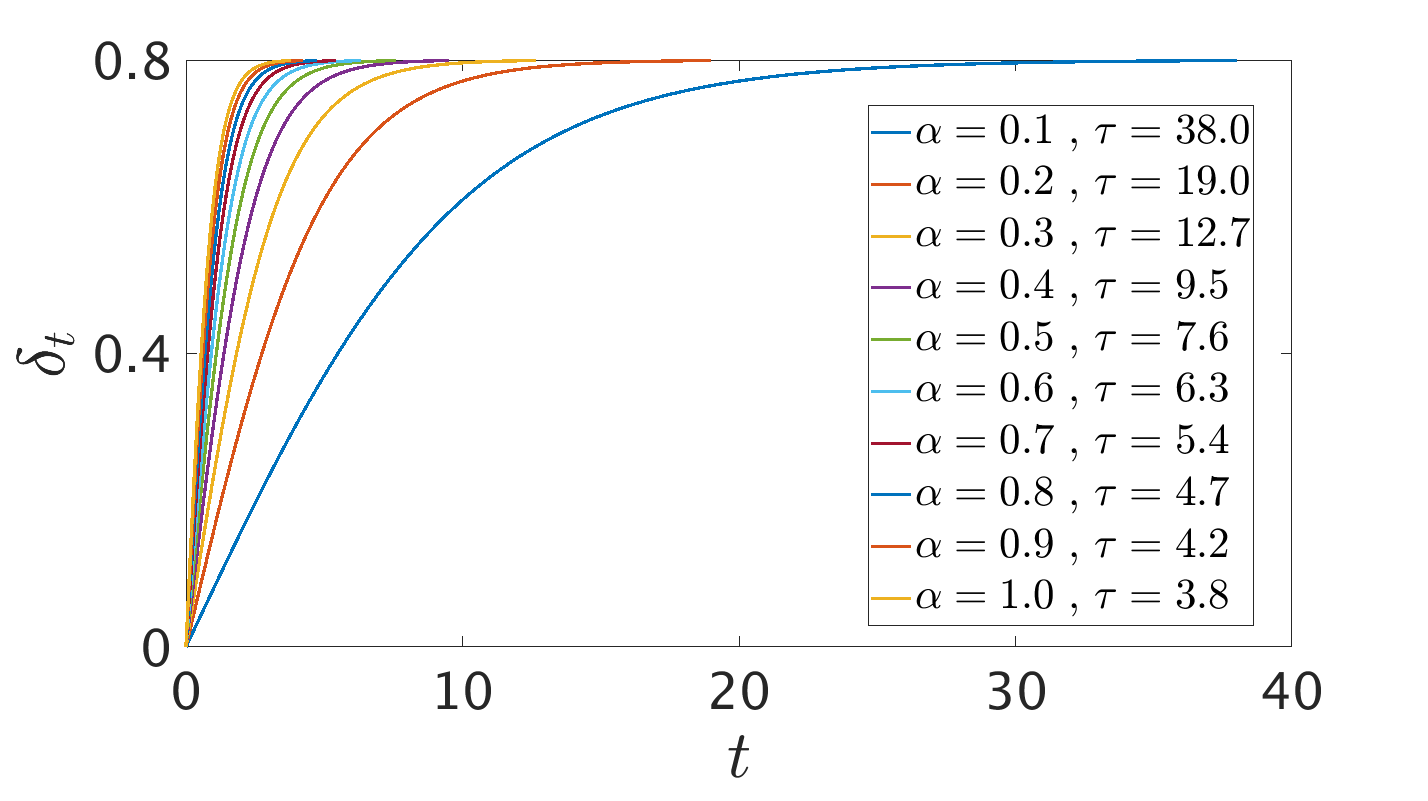}
    \caption{Driven protocol. The figure shows the expansion protocol defined in Eq.~\eqref{magneticfiledtime} for $\delta_1=0$ and $\delta_2=0.8$, for several values of the driving parameter $\alpha$. The duration of each protocol, indicated in the legend, is chosen such that all protocols connect the same initial and final Hamiltonians within the prescribed numerical tolerance discussed in the text. This construction enables a fair comparison between different driving rates by ensuring that the thermodynamic cycle always starts from $H_{\delta_1}$ and ends at $H_{\delta_2}$. Since we set $\gamma_x=1$, time is dimensionless, as are the parameters $\delta$ and $\alpha$. The corresponding compression protocol follows the same trajectory in parameter space in the opposite direction and is therefore not shown. For all plots we set $\epsilon=0.001$.}
    \label{fig:protocol}
\end{figure}

We start by considering the driving protocol defined in Eq.~\eqref{magneticfiledtime}. Figure~\ref{fig:protocol} shows the expansion stroke, $\delta_t$, as a function of time for different values of the driving parameter $\alpha$. For each value of $\alpha$, the duration of the stroke is determined according to the procedure described above. The figure illustrates how increasing $\alpha$ continuously transforms the driving from a smooth finite-time process into an increasingly quench-like protocol, while preserving the same boundary conditions. The corresponding duration of each stroke is also indicated in the figure. Since the compression stroke is obtained by reversing the expansion protocol, it exhibits the same qualitative behavior and is therefore not shown separately.

\begin{figure}
    \centering
    \includegraphics[width=1\linewidth]{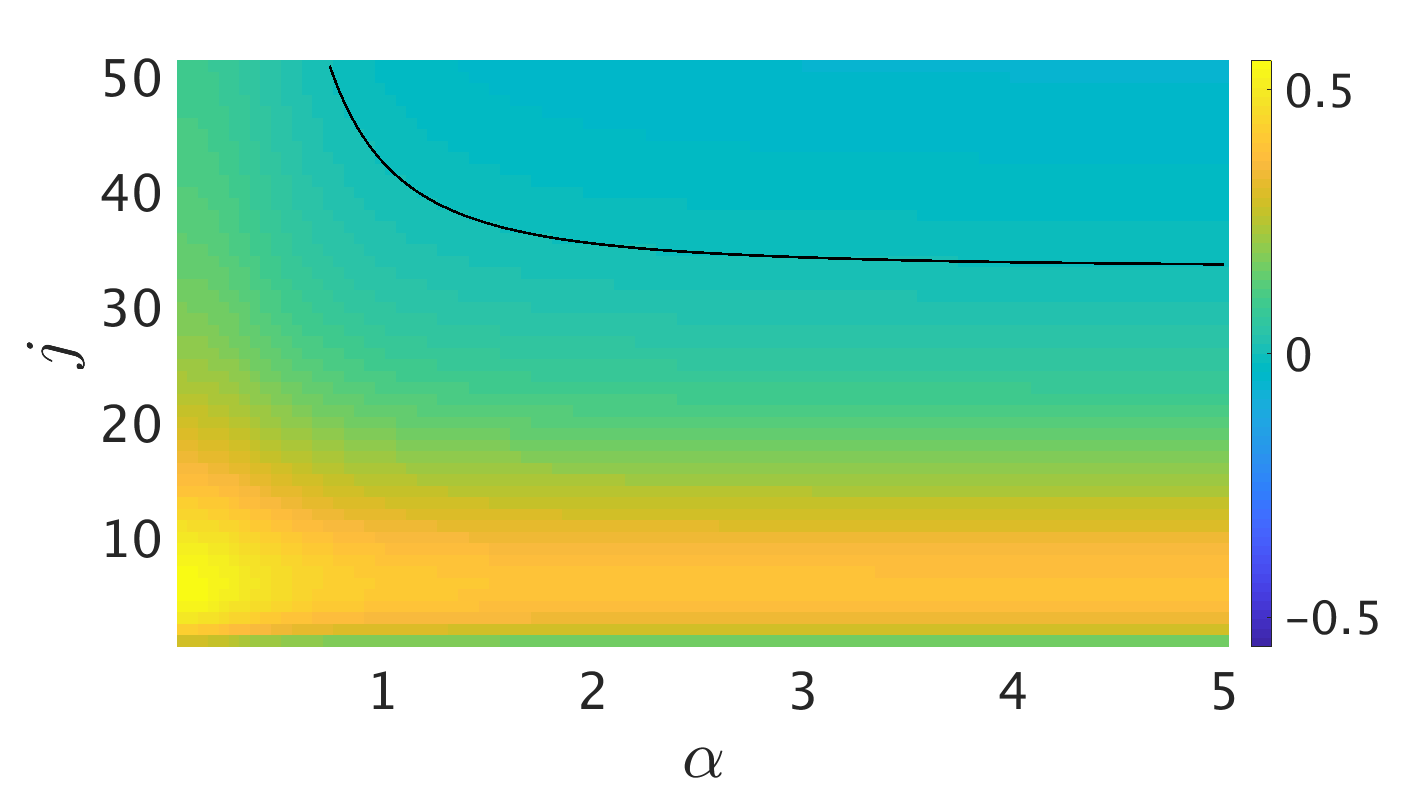}\\
    \includegraphics[width=1\linewidth]{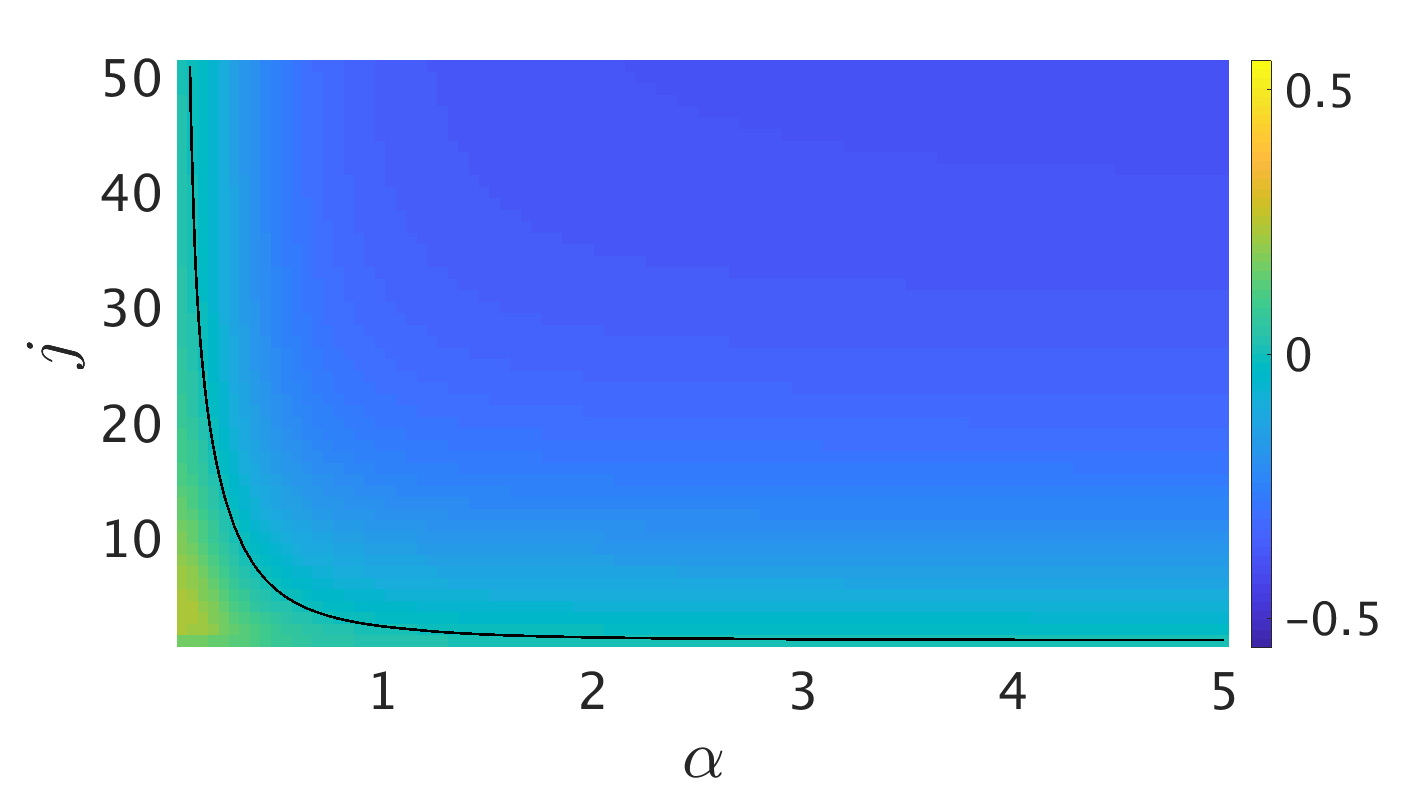}
    \caption{Heat absorbed from the hot reservoir, normalized by $j\delta_0$, as a function of the driving parameter $\alpha$ and the total angular momentum $j$. The upper panel corresponds to $\delta_1=0$ and $\delta_2=0.3$, for which the protocol remains within the broken phase of the LMG model. The lower panel shows the case $\delta_1=0$ and $\delta_2=0.8$, where the protocol crosses the dynamical critical point. The condition $Q_\mathrm{h}=0$, which separates the heat-engine regime ($Q_\mathrm{h}>0$) from a regime in which the hot reservoir no longer supplies energy to the working medium is shown with the black line. Since we set $\gamma_x=1$, the parameters $\delta$ and $\alpha$ are dimensionless. For both plots we take $T_\mathrm{h} = 4T_\mathrm{c}$.}
    \label{fig:heat_env}
\end{figure}

The next thing we need to do is identify the operating range of the engine. Figure~\ref{fig:heat_env} shows the heat absorbed from the hot reservoir. The figure reveals the existence of a critical driving speed for each value of the total angular momentum $j$ beyond which the cycle ceases to operate as a heat engine. At this point, the heat exchanged with the hot reservoir changes sign and the condition $Q_{\mathrm u}^{h}=0$ defines the boundary of the heat-engine regime. The condition $Q_{\mathrm u}^{h}=0$ defines the boundary of the heat-engine regime. When this boundary is crossed, the state $\rho_D$ entering the hot isochoric stroke has a higher average energy, relative to the Hamiltonian $H_{\delta_1}$, than the thermal state $\rho_A$ associated with the hot reservoir. As a consequence, the direction of the energy flux is reversed, and the hot reservoir absorbs energy from the working medium instead of supplying it. From the figure, we see that crossing the critical region substantially reduces the portion of the $(\alpha,j)$ parameter space for which the cycle operates as a heat engine. The operational region defined by the condition $Q_h>0$ becomes substantially smaller when the protocol crosses the critical point. This observation indicates that crossing the critical region enhances the tendency toward heat-flux reversal, rendering the engine more sensitive to finite-time effects.

It should be emphasized that the results shown in Figs.~\ref{fig:heat_env} correspond to a fixed temperature gradient between the hot and cold reservoirs. Consequently, the location of the boundary separating the heat-engine and non-engine regimes depends on the particular choice of thermal bias. In particular, for a fixed driving protocol and a fixed cold temperature, changing the hot-reservoir temperature shifts the region of parameter space where the engine condition is satisfied. Therefore, the operational boundary identified in the present analysis should not be regarded as a universal property of the working medium alone, but rather as a feature resulting from the interplay between the driving protocol and the thermodynamic bias imposed by the reservoirs.

Based on this fact, we define the ratio
\begin{equation}
    \kappa = \frac{\eta_{\mathrm{inv}}}{\eta_{\mathrm{O}}} = \frac{W_{\mathrm{inv}}^{\mathrm{tot}}}{W_{\mathrm{u}}^{\mathrm{tot}}}.
\end{equation}
For the protocols considered here, the coherent heat is found numerically to be non-negative during both unitary strokes. The combination of thermal initial conditions and the specific driving protocol considered here leads to $Q_{\mathrm c}^{\mathrm{exp}}\geq0$ and $Q_{\mathrm c}^{\mathrm{comp}}\geq0$ throughout the parameter range investigated. Hence $Q_{\mathrm c}^{\mathrm{tot}}\geq0$, and the identity $W_{\mathrm u}^{\mathrm{tot}}=W_{\mathrm{inv}}^{\mathrm{tot}}+Q_{\mathrm c}^{\mathrm{tot}}$ implies $W_{\mathrm{inv}}^{\mathrm{tot}}\leq W_{\mathrm u}^{\mathrm{tot}}$. This is intuitively expected since the initial state of both unitary strokes is a thermal equilibrium state. Within the heat-engine regime, where the relevant work output is positive, this yields $0\leq\kappa\leq1$. For quasi-static processes, we have $Q_\mathrm{c}^{\mathrm{tot}} = 0$ and consequently $\kappa = 1$, while for any finite-time process, $\kappa<1$. 

\begin{figure}
    \centering
    \includegraphics[width=1\linewidth]{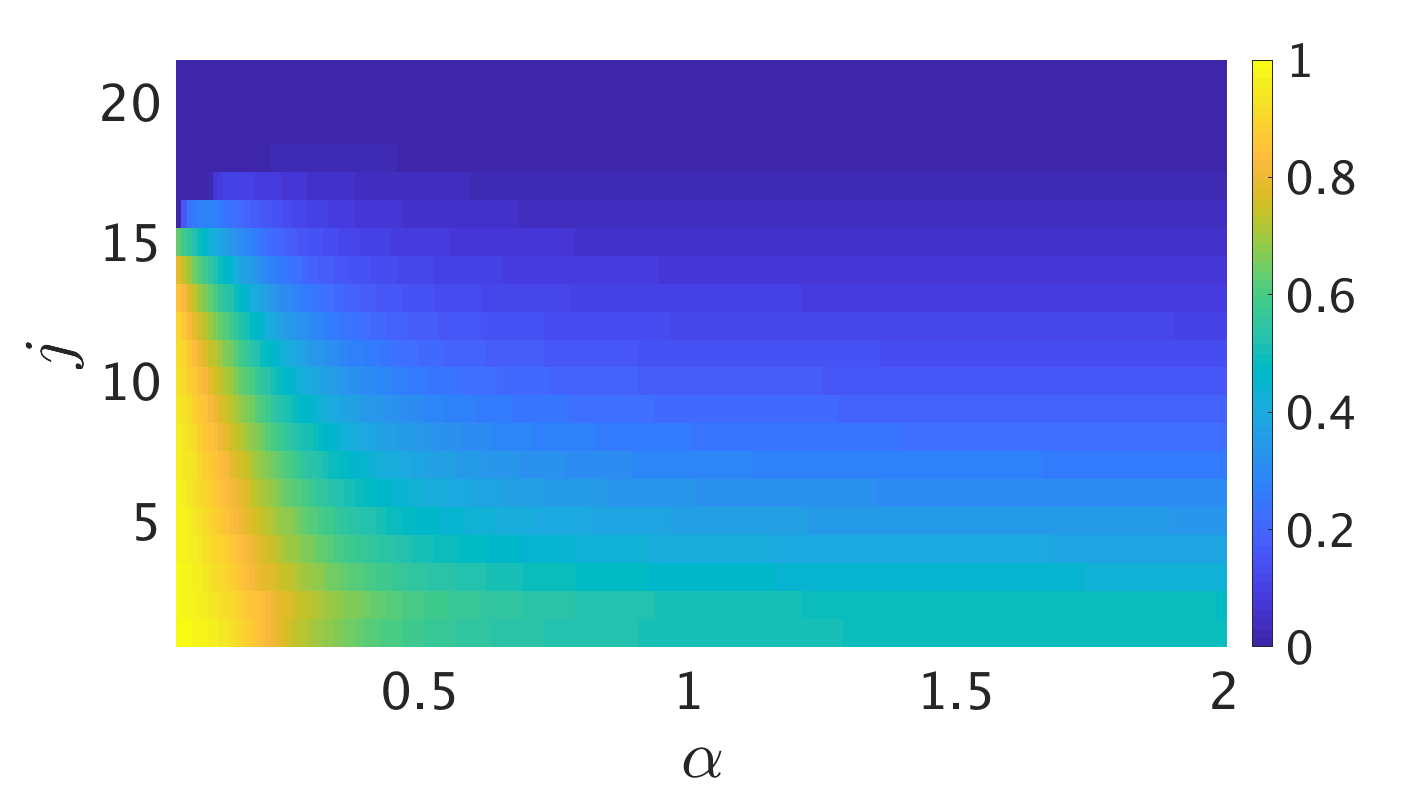}\\
    \includegraphics[width=1\linewidth]{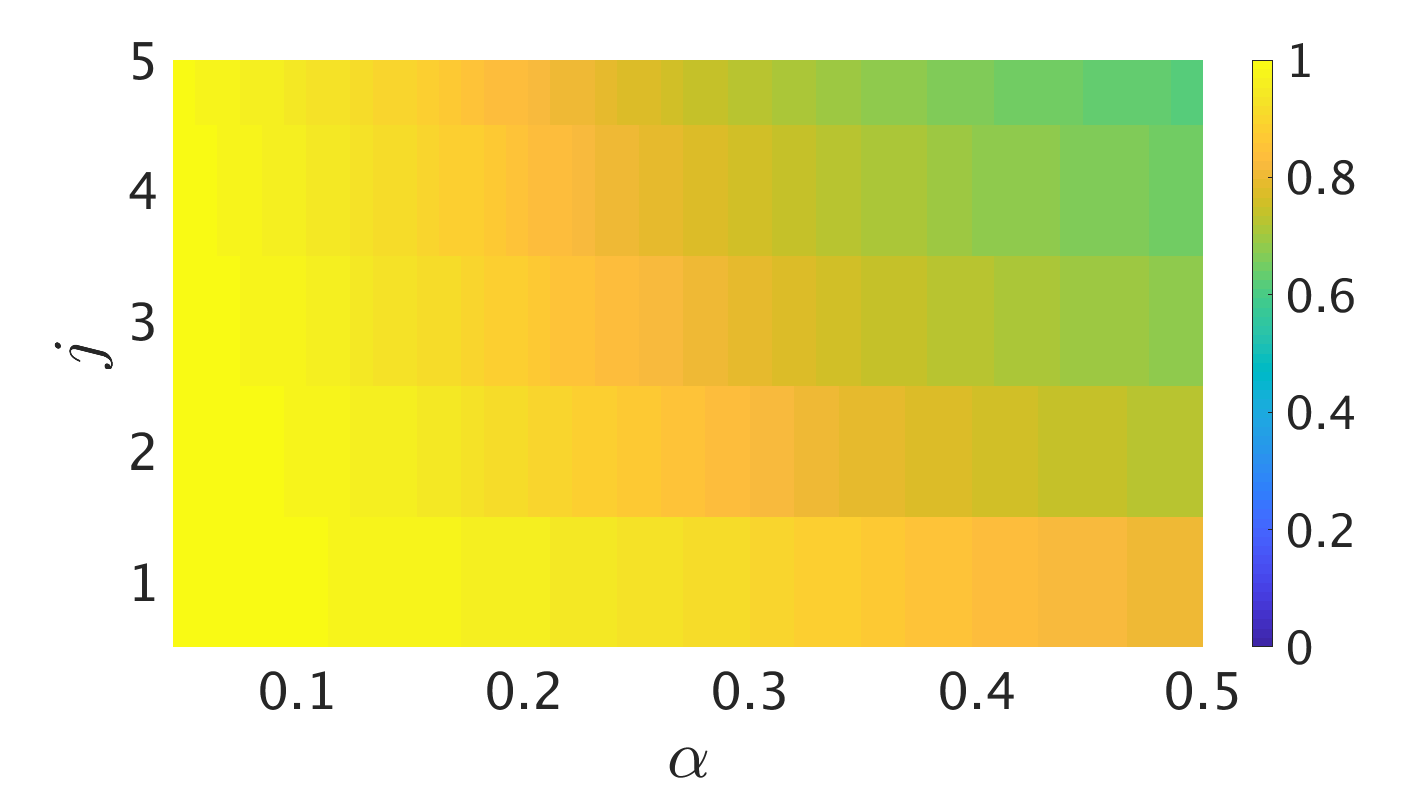}
    \caption{Ratio between the gauge-invariant and the usual work, $\kappa$, shown as a function of the driving parameter $\alpha$ and the total angular momentum $j$. The upper panel corresponds to $\delta_1=0$ and $\delta_2=0.3$, for which the protocol remains within the broken-symmetry phase of the LMG model, while the lower panel shows $\delta_1=0$ and $\delta_2=0.8$, where the protocol crosses the critical point. $\kappa$ measures the fraction of the total work that survives the thermodynamic coarse-graining associated with the gauge-invariant description. Values close to unity indicate that the work is predominantly determined by population dynamics in the instantaneous energy basis, whereas smaller values signal an increasing contribution from the coherent sector. The plots are restricted to the engine's operational region, defined by the condition $Q_h>0$. Since we set $\gamma_x=1$, the parameters $\delta$ and $\alpha$ are dimensionless.}

    \label{fig:kappa}
\end{figure}

Taking into account only the allowed region in the $(\alpha,j)$ plane, we show $\kappa$ in Fig.~\ref{fig:kappa}. Within the operational region, $\kappa$ measures the fraction of the total work that survives the thermodynamic coarse-graining associated with the gauge-invariant description. For $\delta_2=0.3$, where the protocol remains entirely within the broken-symmetry phase, $\kappa$ decreases substantially as $j$ increases, indicating that a significant fraction of the total work is associated with the coherent contribution $Q_c^{\mathrm{tot}}$. In contrast, for $\delta_2=0.8$, where the protocol crosses the critical point and terminates in the non-degenerate phase, one finds $\kappa \approx 1$ throughout most of the operational region. Thus, although the heat-engine regime itself shrinks, the work delivered by the engine becomes predominantly gauge-invariant whenever the cycle remains operational. In this sense, the non-degenerate phase appears to be thermodynamically more effective: once the engine condition is satisfied, most of the extracted work is already contained in the population sector, while only a comparatively small fraction is associated with coherence-induced corrections.

An important consequence of these observations is that the conditions determining whether the cycle operates as a heat engine and the conditions controlling the magnitude of coherent corrections are not the same. Crossing the critical region tends to reduce the domain in which the engine exists, but, whenever the cycle remains operational, it simultaneously suppresses the discrepancy between the usual and gauge-invariant descriptions. This suggests that the mechanisms governing heat-flux reversal and coherent contributions to work are distinct and may be influenced differently by the spectral properties of the LMG model. 

The efficiency decomposition introduced in Eq.~\eqref{eq:split_efficiency} admits an interesting operational interpretation in terms of work extraction under restricted information. To see this, recall that the gauge formulation is based on the idea that not all information contained in the quantum state is thermodynamically accessible. The physically relevant state is instead obtained by averaging over the thermodynamic gauge group~\cite{Celeri2024,Ferrari2025}
\begin{equation}
\rho_{\mathcal G} = \int_{\mathcal{G}_\mathrm{T}}\dd\mathcal{G}\, V\rho V^\dagger,
\end{equation}
where $\dd\mathcal{G}$ denotes the Haar measure on the thermodynamic group $\mathcal{G}_\mathrm{T}$. In the absence of degeneracies, this construction reduces to the dephased state in the instantaneous energy basis, $\rho_{\mathcal G} = \rho^{E,\mathrm{diag}}$. More generally, $\rho_{\mathcal G}$ coincides with the block-diagonal state obtained by averaging over the degenerate energy subspaces.

This observation naturally suggests a restricted notion of ergotropy. Recall that the standard ergotropy is defined as the maximum amount of work that can be extracted from a quantum state through cyclic unitary operations~\cite{Allahverdyan2004},
\begin{equation}
W(\rho,H) = \Tr(\rho H) - \min_U \Tr(U\rho U^\dagger H),
\label{eq:ergotropy}
\end{equation}
where the minimization is performed over all unitary transformations. The resulting quantity measures the maximal work that can be extracted when complete coherent control over the system is available.

However, within the gage formulation, the thermodynamic description is based on the reduced state $\rho_{\mathcal G}$. It is therefore natural to define the gauge-invariant ergotropy as
\begin{equation}
W_{\mathcal G}(\rho,H) = \Tr(\rho_{\mathcal G}H) - \min_U \Tr(U\rho_{\mathcal G}U^\dagger H).
\label{eq:gauge_ergotropy}
\end{equation}
This quantity measures the maximum work extractable from the thermodynamically accessible sector of the state after the information removed by the gauge coarse-graining has been discarded. It is interesting to observe that this quantity is exactly the incoherent contribution to the ergotropy~\eqref{eq:ergotropy} proposed in Ref.~\cite{Francica2020}.

The connection to the invariant work becomes apparent when considering infinitesimal-driven transformations. Since the gauge-reduced state contains only the populations associated with the instantaneous energy levels, the corresponding infinitesimal work extraction is determined exclusively by changes in the spectrum
\begin{equation}
\delta W_{\mathcal G} = \sum_n p^n_t\, \delta\varepsilon^n_t.
\end{equation}
Integrating along the driving protocol, the resulting expression coincides formally with the invariant work of Eq~\eqref{eq:Winv_populations}, suggesting a close connection between invariant work and the gauge-invariant ergotropic sector of the dynamics.

From this perspective, the decomposition $W_{\mathrm u} = W_{\mathrm{inv}} + Q_{\mathrm c}$ acquires a simple operational meaning. The invariant work represents the contribution that remains accessible after the thermodynamic coarse-graining, while the coherent heat quantifies the energetic contribution associated with the coherences removed by that coarse-graining. In the context of the Otto cycle, this interpretation implies that $\eta_{\mathrm{inv}}$ characterizes the efficiency associated with the gauge-invariant ergotropic sector, while $\eta_{\mathrm c}$ measures the modification of the efficiency produced by coherent contributions generated during finite-time unitary strokes.

\subsection{Entropy}

We start by discussing the isochoric strokes. As discussed earlier, no work is performed, and the coherent heat vanishes during these processes. The usual and the invariant approaches lead to the same amount of heat exchanged between the system and the environments. In the quasistatic thermalization limit considered here, Clausius relation states that the total thermodynamic entropy produced in both isochoric processes is given by
\begin{equation}
\Delta S_{\mathrm{th}} =\frac{Q_{\mathrm{inv}}^{\mathrm c}}{T_c} + \frac{Q_{\mathrm{inv}}^{\mathrm h}}{T_h}.
\label{eq:total_iso_th_entropy}
\end{equation}
Therefore, both formulations give the same entropy production. This is expected since the gauge formulation is fully compatible with Classical thermodynamics~\cite{Celeri2024}.

Let us now focus on the unitary strokes of the Otto cycle. During these processes, the working medium is thermally isolated and evolves unitarily under the action of a time-dependent Hamiltonian. Since the initial states of the expansion and compression strokes are thermal states, finite-time driving takes the system out of the corresponding equilibrium manifolds, thereby giving rise to entropy production.

The entropy production associated with a finite-time unitary process can be quantified by the quantum relative entropy between the actual final state and the reference equilibrium state, defined in terms of the final Hamiltonian obtained via an ideal quasistatic adiabatic transformation.

For the expansion stroke, the system evolves from the thermal state $\rho_A$, reaching the nonequilibrium state $\rho_B$. Let $\rho_A^{\mathrm{ad}}$ denote the equilibrium state associated with the final Hamiltonian $H_{\delta_2}$, obtained through the reversible adiabatic transformation. The corresponding entropy production is then given by
\begin{equation}
\Sigma^{\mathrm{exp}} = D\left(\rho_B\,\big\|\,\rho_A^{\mathrm{ad}}\right).
\label{eq:sigma_exp}
\end{equation}

As shown in Ref.~\cite{Plastina2014}, this entropy production is directly related to the inner friction,
\begin{equation}
\Sigma^{\mathrm{exp}} = \beta_A W_{\mathrm{fric}}^{\mathrm{exp}},
\label{eq:sigma_friction_exp}
\end{equation}
where $W_{\mathrm{fric}}^{\mathrm{exp}} = W_{\mathrm u}^{\mathrm{exp}} - W_{\mathrm{ad}}^{\mathrm{exp}}$ and $W_{\mathrm{ad}}^{\mathrm{exp}}$ is the work performed in the corresponding quasistatic adiabatic process. Since $W_{\mathrm u} = W_{\mathrm{inv}} + Q_{\mathrm c}$,
one immediately obtains $W_{\mathrm{fric}}^{\mathrm{exp}} = W_{\mathrm{inv}}^{\mathrm{exp}} - W_{\mathrm{ad}}^{\mathrm{exp}} + Q_{\mathrm c}^{\mathrm{exp}}$. Equation~\eqref{eq:sigma_friction_exp} takes the form
\begin{equation}
\Sigma^{\mathrm{exp}} = \beta_A \left(W_{\mathrm{inv}}^{\mathrm{exp}}-W_{\mathrm{ad}}^{\mathrm{exp}}\right)+\beta_A Q_{\mathrm c}^{\mathrm{exp}}.
\label{eq:sigma_decomposition_exp}
\end{equation}

An analogous reasoning applies to the compression stroke. From the thermal state $\rho_C$, the unitary evolution in finite time produces the nonequilibrium state $\rho_D$. Let $\rho_C^{\mathrm{ad}}$ denote the equilibrium state associated with the final Hamiltonian $H_{\delta_1}$ obtained through the reversible adiabatic transformation. The entropy production is then
\begin{equation}
\Sigma^{\mathrm{comp}} = \beta_C\left(W_{\mathrm{inv}}^{\mathrm{comp}} - W_{\mathrm{ad}}^{\mathrm{comp}}\right) + \beta_C Q_{\mathrm c}^{\mathrm{comp}}.
\label{eq:sigma_decomposition_comp}
\end{equation}

Equations~\eqref{eq:sigma_decomposition_exp} and
\eqref{eq:sigma_decomposition_comp} provides a natural decomposition of the entropy production generated by finite-time driving. The first term measures the contribution associated with changes in the populations of the instantaneous energy eigenstates and survives the thermodynamic coarse-graining. The second term is entirely determined by coherent heat and quantifies the contribution arising from thermodynamically inaccessible coherences generated during the unitary evolution.

Therefore, the entropy production associated with unitary processes admits the decomposition
\begin{equation}
\Sigma = \Sigma_{\mathrm{pop}} + \Sigma_{\mathrm{coh}},
\label{eq:thermo_entropy}
\end{equation}
where
\begin{equation}
\Sigma_{\mathrm{pop}} = \beta \left(W_{\mathrm{inv}} - W_{\mathrm{ad}}\right), \qquad \Sigma_{\mathrm{coh}} = \beta Q_{\mathrm c}.
\end{equation}

Equations~\eqref{eq:sigma_decomposition_exp} and
\eqref{eq:sigma_decomposition_comp} provides a natural decomposition of the entropy generated by the finite-time driving. The first contribution is associated with population redistribution in the instantaneous energy basis and survives the thermodynamic coarse-graining. The second contribution is proportional to the coherent heat and originates from the coherences generated during the driven evolution. In this sense, the gauge formulation resolves the conventional notion of inner friction into a population sector and a coherence sector, allowing one to identify the distinct physical mechanisms responsible for departures from the quasistatic limit.

An interesting consequence of Eqs.~\eqref{eq:sigma_decomposition_exp} and~\eqref{eq:sigma_decomposition_comp} is that the same coherent contribution responsible for the difference between the usual and invariant work also appears in the decomposition of the entropy generated during finite-time driving. Indeed, coherent heat enters simultaneously the energetic relation $W_{\mathrm u}=W_{\mathrm{inv}}+Q_{\mathrm c}$ and the decomposition of the entropy $\Sigma=\Sigma_{\mathrm{pop}}+\Sigma_{\mathrm{coh}}$. The gauge formulation therefore establishes a direct connection between energetic and entropic manifestations of quantum coherence, showing that the quantity responsible for the modification of the work balance also contributes to the departure from the quasistatic limit.

\section{Conclusions}
\label{sec:conclusions}

In this work, we investigated the thermodynamics of a quantum Otto engine from the perspective of gauge-invariant quantum thermodynamics. Taking the Lipkin-Meshkov-Glick model as the working medium, we analyzed the finite-time operation of the cycle and compared the conventional thermodynamic description with the gauge-invariant formulation. The latter is based on the idea that thermodynamics has access only to a restricted subset of the information contained in the quantum state, naturally leading to a separation between population and coherence contributions. Within this framework, the work performed during the unitary strokes decomposes into an invariant contribution, associated with changes in the energy spectrum and populations in the instantaneous energy basis, and a coherent contribution generated by finite-time driving. This decomposition induces a corresponding splitting of the engine efficiency and provides a natural way to quantify the thermodynamic role of quantum coherences.

Using a finite-time driving protocol that connects the same initial and final Hamiltonians across all process speeds, we identified the engine's operational region under the condition that heat must flow from the hot reservoir to the working medium. We showed that crossing the critical region of the LMG model significantly reduces the portion of parameter space for which the cycle operates as a heat engine. At the same time, the gauge decomposition reveals a nontrivial behavior of the work output. By introducing the ratio $\kappa$ between invariant and usual work, we quantified the fraction of extracted work that survives the thermodynamic coarse-graining. Although protocols remaining in the broken-symmetry phase exhibit substantial coherent corrections for sufficiently large systems, protocols crossing the critical point were found to satisfy $\kappa\approx 1$ throughout most of the operational region. Thus, although criticality, in the present model, reduces the thermodynamic viability of the cycle, it simultaneously suppresses the discrepancy between the usual and gauge-invariant descriptions whenever the engine remains operational.

The gauge formulation also provides a new perspective on finite-time irreversibility. In the conventional treatment of quantum heat engines, inner friction is introduced by comparing the actual work performed during a finite-time process with the work obtained in an ideal quasistatic adiabatic transformation. In the present framework, this separation emerges directly from the gauge decomposition of the first law and from the corresponding decomposition of entropy production. As a result, the conventional notion of inner friction naturally resolves into a population contribution and a coherence contribution. One of the main advantages of the geometric formulation is therefore that it identifies the coherent component of finite-time irreversibility from the structure of the thermodynamic reduction itself, rather than through the introduction of additional phenomenological quantities.

This interpretation is closely related to the extraction of work under restricted information. The gauge construction associates with each state a reduced thermodynamic description containing only the information that survives the thermodynamic coarse-graining. From this perspective, invariant work can be interpreted as the contribution accessible within the reduced description, while $\kappa$ quantifies the fraction of total engine performance that can be attributed to this thermodynamically accessible sector. The complementary contribution is associated with the coherence discarded by the gauge reduction. Consequently, the gauge efficiency should not be interpreted as the efficiency of a different engine, but rather as the contribution of the gauge-invariant sector to the efficiency of the same physical cycle.

More broadly, the present results illustrate the role of geometry in quantum thermodynamics. The thermodynamic gauge group specifies which information is retained and which information is removed by the thermodynamic description. Invariant work, coherent heat, efficiency decomposition, and entropy-production splitting all arise as consequences of this geometric structure. In this sense, the gauge formalism does not merely reformulate conventional thermodynamics. Rather, it provides a systematic framework for separating and quantifying the population and coherence sectors of finite-time quantum thermodynamic processes.

Several open questions naturally emerge from the present results. The first direction concerns the role of the spectral structure of the working medium. Our numerical analysis shows that crossing the critical region substantially reduces the operational domain of the engine while simultaneously driving $\kappa$ toward unity. Understanding the microscopic origin of this behavior remains an open problem. In particular, it would be interesting to determine whether the suppression of coherent contributions observed in the non-degenerate phase is related to changes in degeneracies, avoided level crossings, or more general spectral properties of the instantaneous Hamiltonian.

The second direction concerns the coherent contribution itself. Throughout the operational region explored in this work, the coherent heat was found numerically to be non-negative during both unitary strokes. Although the gauge formalism does not impose any sign constraint on this quantity, this observation suggests that additional physical mechanisms may be at play in the LMG model under the driving protocols considered here. Establishing rigorous conditions under which coherent contributions acquire a definite sign and clarifying their relation to nonadiabatic transitions, quantum criticality, and finite-size effects would deepen our understanding of the energetic role of quantum coherences in finite-time thermodynamics.

The present results also motivate a more systematic investigation of gauge-invariant ergotropy. The interpretation proposed here suggests that the gauge-reduced state naturally defines a restricted work-extraction problem in which only the thermodynamically accessible sector contributes to the available work. Developing this idea into a complete resource-theoretic framework could provide a precise operational meaning to invariant work and clarify its relation to conventional notions of ergotropy, passivity, and work extraction under restricted control.

Finally, it would be important to determine the extent to which the phenomena reported here are specific to the LMG model or reflect more general features of finite-time quantum thermal machines. Applying the gauge-invariant formulation to other many-body systems, particularly models exhibiting quantum phase transitions, excited-state criticality, or chaotic dynamics, may help establish whether the separation between population and coherence sectors uncovered in this work constitutes a generic feature of nonequilibrium quantum thermodynamics.

\begin{acknowledgments}
MBS acknowledges support from Brazilian Federal Agency for Support and Evaluation of Graduate Education (CAPES). LCC acknowledges support from the National Council for Scientific and Technological Development (CNPq) through grant 308065/2022-0, the National Institute of Science and Technology for Applied Quantum Computing through CNPq grant 408884/2024-0, Goiás State Research Foundation (FAPEG) through grant 202510267001843, and São Paulo State Research Foundation (FAPESP) through grant 2025/23726-4. TRO acknowledges support from the National Council for Scientific and Technological Development (CNPq) through grant 310682/2023-1, the National Institute of Science and Technology for Applied Quantum Computing through CNPq grant 408884/2024-0 and funding from the Air Force Office of Scientific Research under Grant No. FA9550-23-1-0092.
\end{acknowledgments}


\end{document}